\documentclass[12pt]{article}
\begin{document}
\title{Medium influence of the nucleon nucleon cross section on the fragmentation} 
\author{\small Jian-Ye Liu(刘建业)$^{1,2,3,4}$, Yong-Zhong Xing(邢永忠) $^{1,2}$, 
Wen-Jun Guo(郭文军) $^{3}$} 
\date{}
\maketitle
\begin{center}
$^{1}${\small Institute for the theory of modern physics, Tianshui Normal University, Gansu,
Tianshui 741000, P. R. China}\\
$^{2}${\small Center of Theoretical Nuclear Physics, National Laboratory of 
Heavy Ion Accelerator}\\
{\small Lanzhou 730000, P.R. China}\\
$^{3}${\small Institute of Modern Physics, Chinese Academy of Sciences, P.O.Box 31}\\
{\small Lanzhou 730000, P.R. China}\\
$^{4}${\small CCAST(Word Lab.),P.O.Box 8730,Beijing 100080}\\
{\small Lanzhou 730000, P.R. China}\\
\date{}
\maketitle
\begin{minipage}{140mm}
\baselineskip 0.2in
\hskip 0.2in Based on an isospin dependent quantum
molecular dynamics model we studied the influence of a medium correction of
an isospin dependent nucleon nucleon cross section on the
fragmentation at the intermediate energy heavy ion collisions. We found that the
medium correction from an isospin dependent nucleon nucleon cross
section increases the dependence of the fragmentation on the isospin effect of 
in-medium nucleon nucleon cross section, at the same time, the momentum dependent interaction
(MDI) produces also an important role for enhancing the influence of the medium correction
on the isospin effect of two-body collisions in the fragmentation process.\\\  
{\bf PACS number(s)}: 25$\cdot$70$\cdot$pq\\
Correspond auther: E-mail address: Liujy@ns.lzb.ac.cn\\
Phone number:86-0931-4969318(O),8272215(H). Fax number:86-0931-4969201
\end{minipage}
\end{center}
\baselineskip 0.3in
\hskip 0.3in The rapid progress in producing energetic radioactive
beams has offered an excellent opportunity to investigate various
isospin effects in the dynamics of nuclear reaction ${[1-5]}$.
This study has made it possible to obtain crucial information
about the equation of state of isospin asymmetric nuclear matter
and the isospin dependent in-medium nucleon nucleon cross
section. Recently, several interesting isospin effects in
heavy ion collisions have been explored both experimentally
${[6-16,36]}$ and theoretically ${[17-29,37,38]}$. For a recent
review, see, e.g., Ref. [1]. However, two essential ingredients in
heavy-ion collision dynamics, i.e., the symmetry potential of the
mean field and the isospin dependent in-medium nucleon nucleon
cross section are still not well determined.
Based on an isospin dependent quantum molecular dynamics model(IQMD) we
studied the underlying influence of the isospin effect of an
in-medium nucleon nucleon cross section on the fragmentation and
the dissipation in intermediate energy heavy ion collisions. We
found the pronounced isospin effects in the nuclear stopping, the
multiplicity of intermediate mass fragments and the
number of nucleon emissions due to mainly the isospin dependent in-medium nucleon 
nucleon cross section ${[35,36,37]}$. Furthermore, we investigate the
influences of the medium correction of an isospin dependent nucleon
nucleon cross section on the isospin effects of two-body collision in fragmentation at 
intermediate energy heavy ion collisions. The calculated results show that the medium 
correction of an isospin dependent nucleon nucleon cross section enhances the dependence of the
multiplicity of intermediate mass fragments $N_{imf}$ on the isospin effect of in-medium nucleon
nucleon cross sections, while MDI produces an important role for inducing above effects in the 
fragmentation process.
\hskip 0.3in
As well know that quantum molecular dynamics (QMD) model [30,31] contains two dynamical 
ingredients,the density dependent mean field and the in-medium nucleon nucleon cross section. 
In order to describe the isospin dependence appropriately, the QMD model should be modified 
properly. Considering the isospin effects in mean field,two-body collision and Pauli 
blocking as shown in Ref.[38] we have made important modifications in  QMD to obtain an
isospin dependent quantum molecular dynamics (IQMD) $[1,34]$. The initial density distributions
of the colliding nuclei in IQMD are obtained from the calculations of the Skyrme-Hatree-Fock
with parameter set SKM$^{*}$ $[32]$. The parameters of interaction potential was obtained
by comparing the calculated results with the experimental data of the ground state properties
of coliiding nuclei. The interaction potential is
\begin{equation}
U(\rho)=U^{Sky}+U^{c}+U^{sym}+U^{Yuk}+U^{MDI}+U^{Pauli},
\end{equation}
where  $U^{Sky}$, $U^{c}$, $U^{Yuk}$ , $U^{MDI}$  and $U^{Pauli}$ are Skyrme potential
, Coulomb potential, the Yukawa potential, the momentum dependent interaction and Pauli
potentials. We used the following two different forms of the symmetry potentials 
${[1,2]}$
\begin{equation}
U_1^{sym}=cu\delta \tau _z   ,
\end{equation}
\begin{equation}
U_2^{sym}=\pm
cu^{1/2}\delta-\frac{1}{4}cu^{1/2}\delta^{2}   ,
\end{equation}
\begin{equation}
U_0^{sym}=0.0
\end{equation}
 where
    \[\tau_{z}=\left\{ \begin{array}{ll}
              1 & \mbox{for neutron}\\
             -1 & \mbox{for proton},
             \end{array}
            \right . \]\
and c is the strength of the symmetry potential being a value of 32MeV and $U_0^{sym}$
means without symmetry potential. In this work,
$u= \frac \rho {\rho_0}$ is the reduced density and
$\delta =\frac{\rho _n-\rho _p}{\rho _n+\rho _p}=\frac{\rho _n-\rho _p}\rho $  is
 the relative neutron excess.
$\rho $, $\rho _{_0}$, $\rho _n$ and $\rho _p$ are the total ,
normal, neutron and proton densities, respectively. 
The recent studies of collective flow
in heavy ion collisions at intermediate energy have indicated a
reduction of in-medium nucleon nucleon cross sections. An
empirical density dependent nucleon nucleon cross section in
medium $[26]$ has been suggested as follows
\begin{equation}
\sigma_{NN}=(1+\alpha\frac{\rho}{\rho_{0}})\sigma^{free}_{NN} ,
\end{equation}
 where $\sigma^{free}_{NN}$ is the experimental nucleon nucleon cross section $[33]$.
The expression (5) with  $\alpha$ $ = -0.2$ has been found
to reproduce the flow data. The free neutron proton cross
section is about a factor of 3 times larger than the free proton
proton or the free neutron neutron one below 400 MeV, which
contributes the main isospin effect from nucleon-nucleon
collisions.  The main physics ingredients and their numerical realization 
in the IQMD model can be found in Refs $[1,30,31,34]$.
\par
\hskip 0.3in The isospin effect of the in-medium nucleon nucleon
cross section on the observables is defined by the difference
between the observable from an isospin dependent nucleon nucleon
cross section  $\sigma^{iso}$ and that from an isospin independent
nucleon nucleon cross section $\sigma^{noiso}$ in the medium. Here
$\sigma^{iso}$ is defined as $\sigma_{np} \geq
\sigma_{nn}$=$\sigma_{pp}$ and $\sigma^{noiso}$ means
$\sigma_{np}$ = $\sigma_{nn}$ = $\sigma_{pp}$, where
$\sigma_{np}$, $\sigma_{nn}$ and $\sigma_{pp}$ are the neutron
proton, neutron neutron and proton proton cross sections in medium
respectively.
In order to study the influence played by the medium correction of the
isospin dependent nucleon nucleon cross section on the isospin
effects of two-body collision in the fragmentation process including the contributions from all 
of impact parameters to multiplicity of intermediate mass fragments $<N_{imf}>_{b}$ ,
Fig.1 shows the impact parameter average values of $<N_{imf}>_{b}$  
(from at equilibrium time $ 200 fm/c $) as a function of the beam energy
for the mass symmetry system $^{76}K_{r}+^{76}K_{r}$ (top panels) and mass asymmetry system 
$^{112}S_{n}+^{40}C_{a}$ (bottom panels) with
the same system mass for the different symmetry potentials
$U_1^{sym}$ , $U_2^{sym}$ and $U_0^{sym}$ ,with the isospinm dependent in-medium nucleon 
nucleon cross section $\sigma^{iso}$ and the isospin independent one
$\sigma^{noiso}$. Namely,there are $U_0^{sym}$+$\sigma^{iso}$ 
,$U_1^{sym}$+$\sigma^{iso}$ ,$U_2^{sym}$+$\sigma^{iso}$ ,
$U_1^{sym}$+$\sigma^{noiso}$  and $U_2^{sym}$+$\sigma^{noiso}$  with  
$\alpha =-0.2$ (left panels) and $\alpha = 0.0$ (right panels) in Fig.1. 
The multiplicity of the intermediate mass fragments $<N_{imf}>_{b}$
is defined as the fragments with charge numbers from 3 to 13.
It is clear to see that
all of lines with filled symbols are larger than those with open symbols,i.e.,all of 
$<N_{imf}>_{b}$  with $\sigma^{iso}$
are larger than those with $\sigma^{noiso}$ because the collision number with $\sigma^{iso}$
is larger than that with $\sigma^{noiso}$.We also found that the gaps between  
lines with the filled symbols and the open symbols are larger but the variations 
among lines in each group are smaller.
These mean that $<N_{imf}>_{b}$  depends sensitively on the isospin effect of in-medium 
nucleon nucleon cross section and weakly on the symmetry potential. In particular,
the gaps between two group lines with $\alpha =-0.2$ in left panels are larger than
corresponding those with $\alpha = 0.0$ in right panels, i.e., the medium correction
of two-body collision enhances the dependence of $<N_{imf}>_{b}$ to its isospin effect.
We also found an important role of the MDI for enhancing the
 isospin effect correlating with the medium correction of two-body collision on the 
$<N_{imf}>_{b}$ . Fig.2 shows the time evolusion of $N_{imf}$ for a symmetry potentials
$U_1^{sym}$ at beam energy of 100 MeV/nucleon and impact parameter of 4 fm. They are four cases:
(1) $\alpha =-0.2$ +$\sigma^{iso}$ ,(2) $\alpha = 0.0$ +$\sigma^{iso}$ 
,(3) $\alpha =-0.2$ +$\sigma^{noiso}$  and (4) $\alpha = 0.0$ +$\sigma^{noiso}$ with MDI
in the left window and NOMDI in the right window. It is clear to see that
the gap between the solid line and dot line with MDI in the left window are larger than 
corresponding gap with NOMDI in the right window in the medium($\alpha = 0.-2$), 
i.e., MDI increases the isospin effect of two-body collision on the $N_{imf}$ in the medium 
because above gap is produced from the isospin effect of nucleon-nucleon cross section in the 
medium. Why does the medium correction of nucleon nucleon cross section and MDI enhance the dependence
of $N_{imf}$ on the isospin effect of two-body collision?  
Physically there are three mechanisms at work here.(1) The average momentum of a 
particle in medium is higher in a heavy ion collision than in cold nuclear matter at the
same density. (2) MDI induces the transporting momentum more
effectively from one part of the system to another, in which particles also move with a 
higher velocity for a given momentum than in free space. (3) As well know that the 
isospin dependent in-medium nucleon nucleon 
cross section is a sensitive founction of the nuclear density distribution and beam energy
as shown in Eg.(5). Fig.3 shows the time evolution of the ratio of nuclear
density to normal one, $\frac{\rho} {\rho_{0}}$ for four cases:  they are 
$\rho(\sigma^{iso},\alpha = -0.2)$, $\rho(\sigma^{noiso},\alpha = -0.2)$
, $\rho(\sigma^{iso},\alpha = 0.0)$  and $\rho(\sigma^{noiso},\alpha = 0.0)$  for
the reaction $^{76}$Kr + $^{76}$Kr and symmetry potential
$U_1^{sym}$ at E= 150 MeV/nucleon and b= 4.0 fm.
From the values of peak for $\frac{\rho} {\rho_{0}}$ in the insert in Fig.3 it is clear to 
see that  $\rho(\sigma^{iso},\alpha = -0.2)$ (solid line) is larger
than $\rho(\sigma^{noiso},\alpha = -0.2)$ (dot line) and  $\rho(\sigma^{iso},\alpha = 0.0)$
(dashed line)is larger than $\rho(\sigma^{noiso},\alpha = 0.0)$ (dot-dashed line) because
the larger collision number from $\sigma^{iso}$ increases the nuclear stopping and dissipation
, which enhances the nuclear density, compared to the case with $\sigma^{noiso}$. From Fig.3 we 
can also see that $\frac{\rho} {\rho_{0}}$ decrease quickly with increasing the time after peak
of $\frac{\rho} {\rho_{0}}$. Because the 
larger compression produces quick expanding process of the colliding system ,while the small 
compression induces slow expanding process
, at the same time, the $\frac{\rho} {\rho_{0}}$ decrease quickly with expanding process of 
system. But the decreasing velocity of  $\frac{\rho} {\rho_{0}}$ is larger for the quick expansion 
system than that for the slow expansion system, up to about after 70 fm/c, on the contrary, 
$\rho(\sigma^{noiso},\alpha = -0.2)$ (dot line) 
is larger than $\rho(\sigma^{iso},\alpha = -0.2)$ (solid line) and 
$\rho(\sigma^{noiso},\alpha = 0.0)$ (dot-dashed line) is larger than 
$\rho(\sigma^{iso},\alpha = 0.0)$ (dashed line). In particular, the gap between two lines for
$\alpha = -0.2$ is larger than that for $\alpha = 0.0$ after about 70 fm/c. This property is 
very similar to the $N_{imf}$, which means that the medium correction of an isospin
dependent nucleon nucleon cross section enhances also the
dependence of $\rho$ on the isospin effect of two-body collision,including the role of MDI on
 it , which induces the same effects to
$N_{imf}$ through the nucleon nucleon cross section as a function of the nuclear density
as shown in Eg.(5).\\\
\hskip 0.3in 
In summary,the calculation results by using IQMD show that
(1) the multiplicity of intermediate mass
fragments $N_{imf}$ with $\sigma^{iso}$ are always larger than those with
$\sigma^{noiso}$ in the energy region studied here.
(2) In particular,the difference between the two values of $N_{imf}$ from
the $\sigma^{iso}$ and $\sigma^{noiso}$ for $\alpha=-0.2 $
is larger than corresponding that for $\alpha=0.0 $, i.e., the medium
correction of the isospin dependent nucleon nucleon cross section
enhances the dependence of $N_{imf}$ on the isospin
effect of isospin dependent nucleon nucleon cross section .
(3) MDI produces an important role for enhancing the isospin effect of two-body collision 
from the medium correction on the $N_{imf}$.
\section{ACKNOWLEDGMENT}
We thank Prof.Bao-An Li for helpful discussions.\\\
This work is supported by the Major State Basic Research Development Program
in China Under Contract No.G2000077400, "100-person project" of the Chinese
Academy of Sciences, the National Natural Science Foundation of China under Grants
Nos.10175080, 10004012,10175082 and The CAS Knowledge Innovation Project No. KJCX2-SW-N02\\\

\baselineskip 0.2in
\section*{Figure captions}
\begin{description}
 \item[Fig.1]
 An impact parameter average value of the multiplicity of intermediate mass
 fragments,$<N_{imf}>_{b}$ as a function of the beam energy  for  $U_0^{sym}$, $U_1^{sym}$ 
and $U_2^{sym}$ with $\sigma^{iso}$ and $\sigma^{noiso}$ for the mass symmetry system
$^{76}$Kr + $^{76}$Kr  and asymmetry mass system $^{112}$Sn + $^{40}$Ca(see text).
 \item[Fig.2]
 The time evolution of the $ N_{imf}$ with MDI and NOMDI for four cases (see text) and systems
$^{76}Kr+^{76}Kr$ at E=150 MeV/nucleon and b= 4.0 fm.
\item[Fig.3]The time evolution of the ratio of nuclear density to normal density 
 $\frac{\rho(\sigma^{iso},\alpha )} {\rho_{0}}$ for different cases(see text).

\end{description}

\end{document}